\documentclass[twocolumn]{aastex62}

\usepackage{latexsym}
\usepackage{color}
\usepackage{amsmath}
\usepackage{amssymb}
\usepackage{graphicx}
\usepackage{aas_macros}
\usepackage{bm}

\newcommand{\Ms}{M$_{\odot}$}
\newcommand{\Hmol}{H$_2$}
\newcommand{\s}{$\sigma_\mathrm{bc}$}

\newcommand{\Rvir}{{\ensuremath{R_\mathrm{vir}}}}
\newcommand{\Rhalf}{{\ensuremath{R_{1/2}}}}

\newcommand{\cc}{cm$^{-3}$}

\graphicspath{{./}{figures/}}

\submitjournal{ApJ}

\shorttitle{Globular Clusters and Streaming Velocities in high-resolution cosmological simulations}
\shortauthors{Schauer et al.}

\begin{document}

\title{Globular Clusters and Streaming Velocities: Testing the new formation channel in high-resolution cosmological simulations}

\correspondingauthor{Anna T. P. Schauer}
\email{anna.schauer@utexas.edu}

\author[0000-0002-2220-8086]{Anna T. P. Schauer}
\altaffiliation{Hubble Fellow}
\affiliation{Department of Astronomy, 
University of Texas at Austin,
TX 78712, USA}

\author{Volker Bromm}
\affiliation{Department of Astronomy,
University of Texas at Austin,
TX 78712, USA} 

\author{Michael Boylan-Kolchin}
\affiliation{Department of Astronomy,
University of Texas at Austin,
TX 78712, USA} 

\author{Simon C. O. Glover}
\affiliation{Universit\"at Heidelberg, 
Zentrum f\"ur Astronomie, 
Institut f\"ur Theoretische Astrophysik, Albert-Ueberle-Str. 2, 
69120 Heidelberg, Germany}

\author{Ralf S. Klessen}
\affiliation{Universit\"at Heidelberg, 
Zentrum f\"ur Astronomie, 
Institut f\"ur Theoretische Astrophysik, Albert-Ueberle-Str. 2, 
69120 Heidelberg, Germany}
\affiliation{Universit\"{a}t Heidelberg, Interdisziplin\"{a}res Zentrum f\"{u}r Wissenschaftliches Rechnen, Im Neuenheimer Feld 205, D-69120 Heidelberg, Germany}

\begin{abstract}
The formation of globular clusters and their relation to the distribution of dark matter have long puzzled astronomers. One of the most recently-proposed globular cluster formation channels ties ancient star clusters to the large-scale streaming velocity of baryons relative to dark matter in the early Universe. These streaming velocities affect the global infall of baryons into dark matter halos, the high-redshift halo mass function, and the earliest generations of stars. In some cases, streaming velocities may result in dense regions of dark-matter-free gas that becomes Jeans unstable, potentially leading to the formation of compact star clusters.  We investigate this hypothesis using cosmological hydrodynamical simulations that include a full chemical network and the formation and destruction of H$_2$, a process crucial for the formation of the first stars. We find that high-density gas in regions with significant streaming velocities -- which constitute approximately 1\% of the Universe -- is indeed somewhat offset from the centers of dark matter halos, but this offset is typically significantly smaller than the virial radius. Gas outside of dark matter halos never reaches Jeans-unstable densities in our simulations. We postulate that low-level ($Z \approx 10^{-3}\,Z_{\odot}$) metal enrichment by Population III supernovae may enable cooling in the extra-virial regions, allowing gas outside of dark matter halos to cool to the CMB temperature and become Jeans-unstable. Follow-up simulations that include both streaming velocities and metal enrichment by Population III supernovae are needed to understand if streaming velocities provide one path for the formation of globular clusters in the early Universe.
\end{abstract}

\keywords{early universe --- dark ages, reionization, first stars ---
stars: Population III, globular clusters}

\section{Introduction}
Globular clusters (GCs) are old, very dense, and ubiquitous around galaxies. They have been known and studied for hundreds of years, and yet, many aspects of their formation and evolution remain shrouded in mystery. Their nearly universal luminosity function and apparently ancient epoch of formation hints at a common formation mechanism, but a detailed look at their properties indicates a more complicated picture \citep[for reviews of GC formation and properties, see ][]{harris1991,brodie2006,bastian2018,forbes2018}. 

Their relation to dwarf galaxies and connection to the dark matter halos within which all galaxies are thought to form is likewise murky. Individual globular clusters show no evidence for dark matter \citep{moore1996a,conroy2011a,ibata2013a},
yet populations of globular clusters exhibit striking correlations with the inferred dark matter properties of their host galaxies \citep{blakeslee1997,spitler2009,hudson2014}. 
These correlations, coupled with the need for dark matter to allow for baryonic collapse in galaxy formation models constrained by the measured low amplitude of temperature fluctuations in the cosmic microwave background (CMB), has led to many proposals that do inextricably link globular clusters and dark matter.

Historically, this type of model -- which argues that globular clusters formed within massive dark matter minihalos or low-mass atomic cooling halos \citep{peebles84,rosenblatt1988,bromm02,moore2006,trenti15}
-- has been one of two broad classes of GC formation models.  Models connecting GC formation to dark matter are attractive in that they naturally explain the number density, ages, and kinematics of globular clusters, as well as the scaling of properties of galaxies' GC systems with the halo mass. However, the lack of observed dark matter in GCs has cast a persistent shadow over this GC formation picture. On the other hand, we note that the absence of detected dark matter in globular clusters is not {\it prima facie} surprising even in many of these models: for example, no one is troubled about the lack of measured dark matter in giant molecular clouds. 

The other class of models argues that GCs are best understood as the extreme end of the ``normal'' star formation process: giant molecular clouds (GMCs) form with a range of masses and densities, and it may simply be the high-density, high-pressure tail of the GMC distribution that produces globular clusters 
\citep{elmegreen1997,kravtsov2005,el-badry2019}.
This would explain why globular clusters are typically old, as conditions are much more conducive for dense and high-pressure GMCs, but in this picture it is not obvious (1) why the GC luminosity function should look so similar across different environments and (2) whether the age distribution of GCs is compatible with observations.

A completely distinct model for the formation of old, metal-poor globular clusters, invoking the large-scale offset in the bulk velocity of baryons relative to dark matter imprinted at matter-radiation decoupling \citep{th10}, was recently proposed by \citet{naoz14}. 
In this scenario, the so-called baryonic streaming velocity --- discussed in more detail in Section \ref{sec:meth_stream} --- can lead to significant concentrations of gas outside of the virial radii of the earliest-forming dark matter halos \citep{druschke20}, an effect that is not expected if the streaming velocities are ignored. If these clumps become Jeans unstable, they can collapse to form dense stellar clusters that are naturally dark-matter-poor.

This theoretical model of ``Supersonically Induced Gas Objects" (SIGOs) has found recent support in simulations
that include streaming velocities \citep{popa16,chiou18,chiou19,chiou20}.
The SIGOs in these simulations, found outside of dark matter halos' virial
radii, have gas fractions of at least 40\%, well above the cosmic baryon fraction of $\sim 16\%$, at redshifts $z=20-10$.
However,
these simulations only include adiabatic or atomic cooling
and do not follow the collapse of a gas cloud to higher densities.
Molecular hydrogen is crucial to determine if a primordial gas cloud
can cool and form stars \citep{glover13}, and follow-up investigations
are needed to see if these potential proto-globular cluster regions are able to form dense star clusters, and if so, whether those star clusters can survive to the present day. 

\citet{schauer20} presented a set of high-resolution cosmological simulations with a resolution of $\sim$100\,M$_\odot$
in dark matter and of $\sim$20\,M$_\odot$ in gas and explicit treatment of chemical networks that track the formation and destruction of molecular hydrogen, with different amplitudes of streaming velocities. These simulations are therefore ideal to further investigate the scenario of \citet{naoz14} and \citet{chiou18} and determine whether baryon-rich clumps can form outside of dark matter halos, reach the high densities required for globular cluster formation, and subsequently survive for many dynamical times. 

The paper is structured as follows: in Section \ref{sec:meth},
we briefly summarize the physics of streaming velocities and the simulations used for this analysis,
and we introduce the dense region finder used for evaluating dense gas clumps and proto-globular clusters in our simulations.
In Section \ref{sec:stat}, we present our results, starting with a statistical analysis
of the distances of our halos from the gas regions,
followed by characteristics of the high-density gas regions and a Jeans analysis. We also present a heuristic argument about how considering metal-enriched, rather than primordial, gas would alter our analysis. After a discussion in Section \ref{sec:dis}, we summarize our results and point to avenues for future work in Section \ref{sec:conclusions}.
\section{Cosmological context}
\label{sec:meth}
\subsection{Streaming Velocities}
\label{sec:meth_stream}
Streaming velocities are large-scale velocity offsets between baryons and CDM ($\bm{v}_{\rm bc}$) that result from baryonic acoustic oscillations in the photon-baryon plasma prior to recombination. They are identically zero in linear perturbation theory and therefore have long been overlooked. 

However, \citet{th10} noted that this second order effect can be highly significant: streaming velocities are distributed like a three dimensional Gaussian random field with a rms value of $\sigma_{\rm bc} \approx 30$\,km\,s$^{-1}$ at matter-radiation decoupling ($z \sim 1100$). Baryons therefore have significant bulk velocities relative to the dark matter field, and these velocities are a factor of $\sim 5$ larger than the sound speed of baryons immediately after decoupling, making them supersonic. This bulk velocity has a coherence scale of $\sim 3\,{\rm cMpc}$ and its  magnitude $v$ is Maxwell-Boltzmann distributed with a typical value of $v_\mathrm{bc}\approx \sigma_{\rm bc}$ (\citealt{th10}; see \citealt{fialkov14} for a review).

As the Universe expands, the streaming velocities decay as $(1+z)$, rendering them unimportant at low redshifts. At redshifts $z \sim 20-30$, however, they are large enough --- $v_\mathrm{bc} \sim 30\,{\rm km\,s^{-1}}(1+z)/1100\approx 0.5-0.8\,{\rm km\,s^{-1}}$ --- to play a potentially important role in the evolution of early cosmic structure and the formation of Population~III (Pop~III) stars, which are composed of primordial material and are (by definition) metal-free \citep{bromm13}. These stars are expected to form in minihalos, where molecular hydrogen is the only available cooling channel, or in slightly more massive ``atomic cooling" halos, where cooling can proceed via collisional excitation of atomic hydrogen \citep[see e.g.][]{machacek01,yoshida03,hirano15,schauer19,skinner20,schauer20}. The characteristic virial velocity separating the minihalo and atomic cooling halo regimes, assuming a neutral atomic gas with primordial composition, is $11\,{\rm km\,s^{-1}}$, equivalent to a virial temperature of $10^{4}\,{\rm K}$, or a mass of $10^{7}$\,\Ms\ at $z\approx 20$. 
The transition to the next generation of stars, Population~II, happens at an enrichment level of $Z\sim 10^{-3.5}Z_\odot$ \citep{bromm01b,schneider02,klessen2012,glover13}, when gas cooling starts to be influenced by the trace metal fraction. 

The low mass of minihalos and atomic cooling halos implies that the baryon-DM streaming velocity can have a substantial effect on the ability of gas in these halos to cool and eventually form stars. As an example, consider a typical mini-halo with a virial mass of $10^{5}$\,\Ms\ at $z=20$: its virial velocity is approximately $2.4\,{\rm km\,s^{-1}}$, which means that the characteristic streaming velocity of $0.6\,{\rm km\,s^{-1}}$ 
at that time carries baryons with a bulk velocity that is almost $\approx 25\%$ of the virial velocity of a minihalo. 
The impact on the gas primarily occurs prior to virialization, when the escape velocity of the halo is smaller and the streaming velocity is larger.
This makes it significantly more difficult for gas to settle into a minihalo compared to the case where the streaming velocities are absent.  

As gas in streaming velocity regions settles into dark matter halos later, 
the gas content in dark matter halos is reduced \citep{naoz13,schauer19}. Subsequently, high gas densities are reached for more massive halos and the minimum halo mass required for star formation increases \citep{greif11,stacy11,schauer20}. Globally, Pop~III star formation shifts to later times, as more massive halos form at later times. 
\subsection{Simulations}
In this study, we analyze two simulations with 
streaming velocities of 0 and 2\,$\sigma_\mathrm{bc}$  -- labelled 
v0\_lw0 and v2\_lw0 -- 
from the set of simulations by \cite{schauer20}.
In the following, we provide a brief summary of this earlier study.

The simulations were performed with the moving mesh code \textsc{arepo} \citep{arepo}, including dark matter, gas, and sink particles. 
Given a box size of (1\,cMpc/$h$)$^3$ with 1024$^3$ dark matter particles 
and (initial) gas cells, a resolution of $M_\mathrm{DM} \approx 99$\,\Ms\ and $M_\mathrm{gas} \approx 19$\,\Ms\ is achieved throughout the entire box. The runs are initialized at a redshift $z=200$ using \cite{Planck2016} 
cosmological parameters ($h = 0.6774$,  
$\Omega_{\rm m} = 0.3089$, $\Omega_\mathrm{b} = 0.04864$, $\Omega_\Lambda = 1-\Omega_{\rm m}= 0.6911$,  
$n = 0.96$ and $\sigma_8 = 0.8159$). For  v2\_lw0, we include the 2\,\s\,streaming velocity 
at initialization by adding a uniform offset velocity of 12\,km\,s$^{-1}$ 
to the gas cells, chosen (without loss of generality) to be in the $x$-direction. 

The simulations contain a primordial chemistry network \citep{Hartwig15a} to correctly treat the chemical evolution in the high-redshift Universe. 
We follow the species H, H$^+$, \Hmol, D, D$^{+}$, HD, He, He$^{+}$ and e$^{-}$ 
as well as the equilibrium abundances of the H$^{-}$ and H$_{2}^{+}$ ions. Accurately following the formation and destruction of \Hmol\ is of particular importance here, as 
molecular hydrogen is the main coolant for first star formation 
in the absence of metals and dust. 

To study the simplest case and to keep our results comparable to \cite{chiou18}, 
we chose to investigate 
the simulations v0\_lw0 and v2\_lw0 
that do not contain a Lyman-Werner background
(reflected in the simulation name ending in ``lw0'').
Lyman-Werner radiation, emitted by the first stars, 
can destroy H$_2$ and hence hinder star formation \citep{field66,stecher67}, 
moving the star formation threshold to higher halo masses \citep{machacek03}.  
However, the effect of a Lyman-Werner background is small with the inclusion of H$_2$ self-shielding
\citep{skinner20,schauer20}. 
Only a strong Lyman-Werner radiation field, a scenario that that is most probable 
in the immediate vicinity of a star-forming halo, can significantly suppress Pop~III star formation in a minihalo \citep{safranek12}. 
\subsection{Dense gas clumps and proto-globular cluster regions}
We post-process the simulations using both a friends-of-friends algorithm with the default linking length of 0.2 \citep{davis1985} and the subfind \citep{subfind} halo search algorithm. Both algorithms are run on the dark matter structure, while gas is treated as a secondary component that counts towards the total mass of the halo but is not accounted for when identifying the halos. At high redshift, the minihalos and atomic cooling halos identified by the friends-of-friends algorithm tend to be highly aspherical \citep{sasaki14}. The
halos recovered by subfind are more spherical and so we use these as the basis of our study. In the following, we will refer to these subhalos as ``halos'' unless explicitly stated otherwise. Note that each of the halos recovered by subfind is located within a halo identified by the friends-of-friends algorithm.

In our analysis, we consider all halos with masses $M \ge M_{\rm min} = 5\times10^4 \: {\rm M_{\odot}}$. 
These halos are well-resolved (with more than 500 particles / 
gas cells), and our lower mass limit is about an order of  magnitude smaller than 
the minimum halo mass for star formation \citep{schauer20}. 
We consider two radii: 
the virial radius, \Rvir, defined as the radius of a sphere surrounding the halo center which encloses matter with a mean density of 200 times the critical density of the Universe; and the half-mass-radius, \Rhalf, defined as the radius of a sphere that encloses half of the total mass of the halo. \Rvir\ is calculated by the friends-of-friends algorithm, while \Rhalf\ is calculated by the subfind halo finder. 
For concentrations typical of high-redshift minihalos, $0.35  \lesssim \Rhalf/\Rvir \lesssim 0.5$ (assuming a \citealt{navarro1997} density profile).

\begin{figure*}
\centering
\includegraphics[width=2.2\columnwidth]{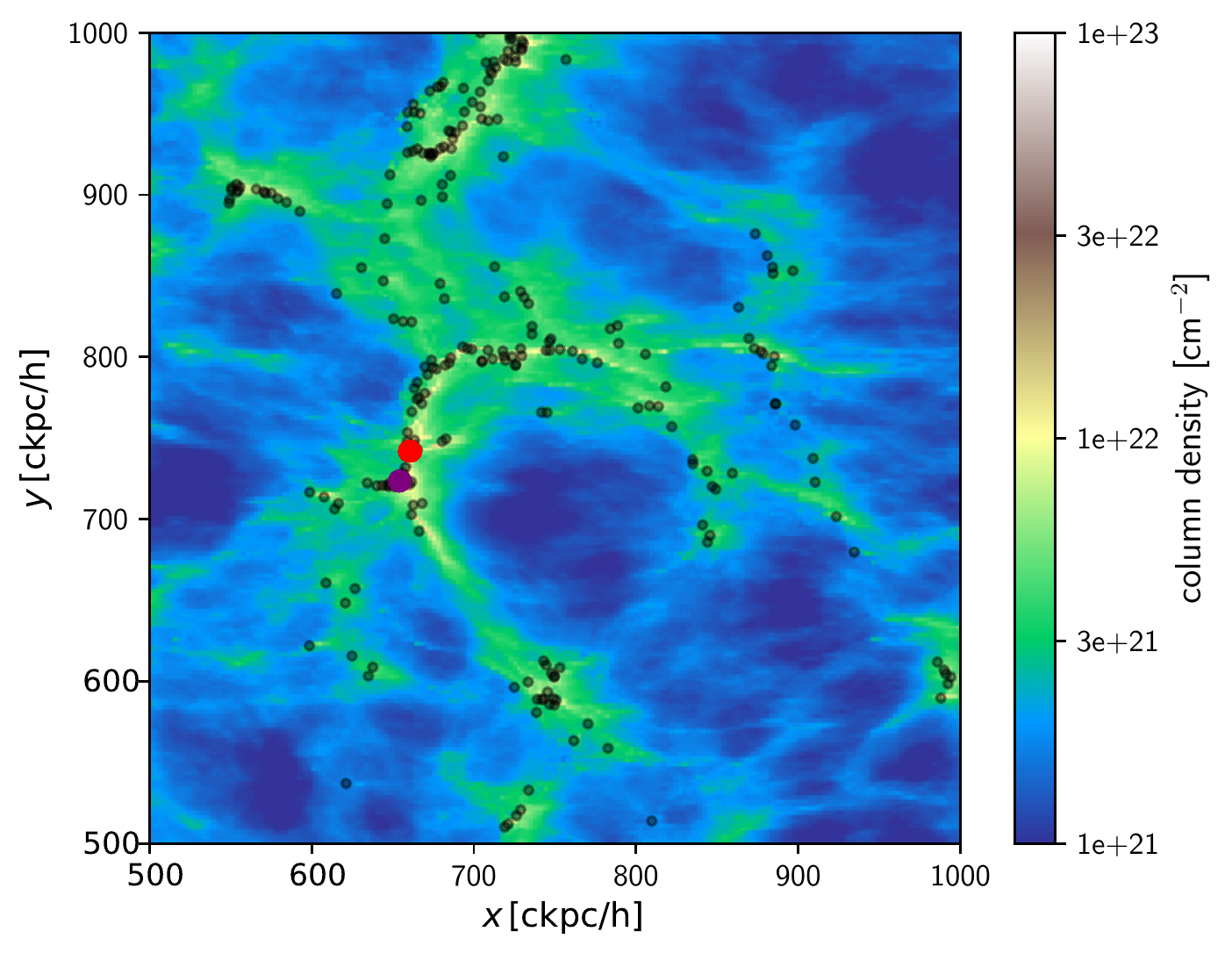}
\caption{Gas column density in a cutout region of the simulation box with a depth of 100\,ckpc/$h$ at redshift $z=15$. All halos with masses greater than $5\times 10^4$\,\Ms\ 
are shown as black dots. The red and purple points indicate the locations of the two regions within this volume that have gas densities $n \ge 100$\,\cc. Almost all of the halos are found along the dense filaments, as are the two highlighted high-density regions. The effect of the streaming velocity is apparent as the gas appears 
``washed-out'' in the $x$-direction. }
\label{fig:overview}
\end{figure*}

Figure~\ref{fig:overview} shows a cutout of the streaming velocity simulation box with a side length of half of the simulation box (500\,ckpc$/h$) and a depth of 100\,ckpc/$h$, which is 10\% of the box size. In the Figure, we show a color-coded map of the gas column densities within this cutout region. The black dots show all of the halos with $M \ge M_{\rm min}$. One can see that almost all of the halos follow the high-density filamentary structure on these 
large scales. As in \cite{schauer20}, one can see the effect of streaming velocities 
on large scales: the gas appears to be ``washed-out'' and elongated 
in the $x$-direction, the direction of the streaming velocity in these simulations. 
We also show in the Figure the locations of the two regions within this cutout that have number densities $n \ge 100$\,\cc\ and that are highly likely to form stars (see Section~\ref{sec:stat} below). These regions are denoted with the red and purple dots. One can see in this large-scale context that these high-density, star-forming regions are closely associated with dark matter halos. 

\begin{figure*}
\centering
\includegraphics[width=0.98\columnwidth]{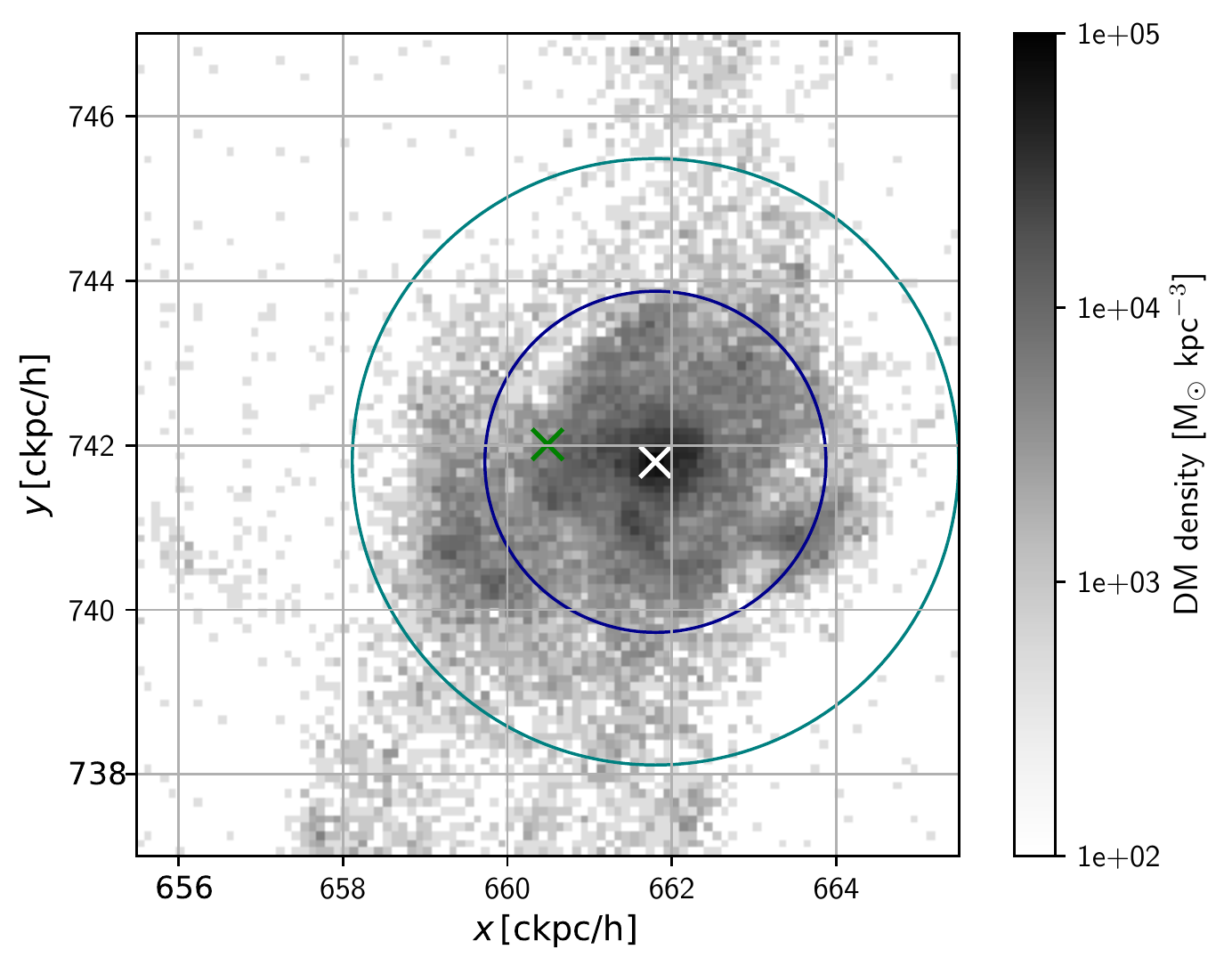}
\includegraphics[width=0.98\columnwidth]{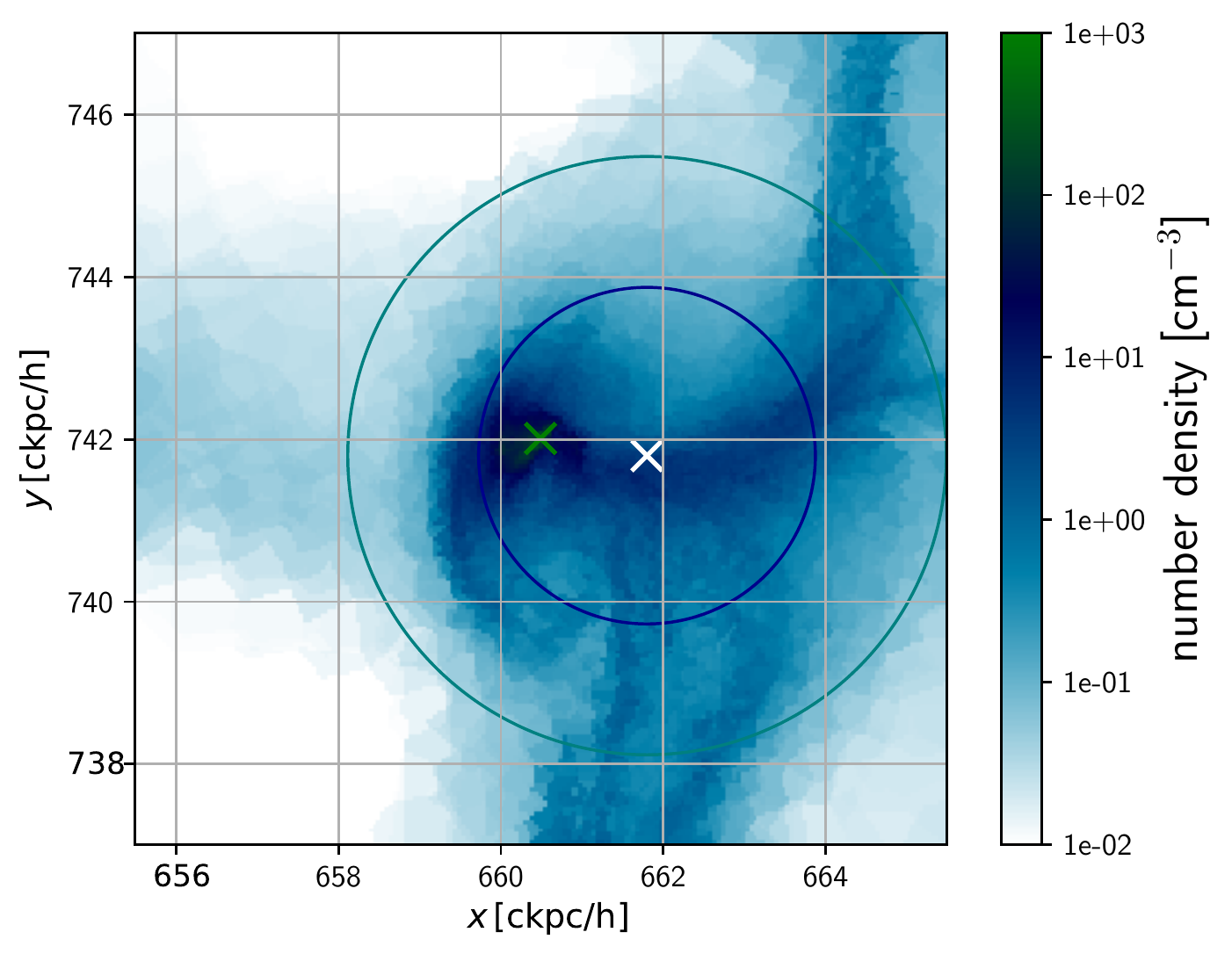}
\caption{Density slice plots of the dense gas clump environment (the purple extended dot in Fig.~\ref{fig:overview}) in the simulation with $2\,\sigma_{bc}$ streaming velocity. Left panel: dark matter density in a 1\,ckpc/$h$ slice. Right panel: gas number density in the same slice. The center of the dark matter halo is shown with a white cross, and its half-mass and virial radii are shown by dark blue and teal circles, respectively. The slices are centered on the cell with the highest gas density, which is shown by a green cross. One can see that the gas region is indeed off-centered, but still within the half-mass radius of the dark matter halo. }
\label{fig:slice}
\end{figure*}

We show a zoom-in into the purple region in Figure~\ref{fig:slice}, as 
this is the gas region with a number density of $n \ge 100$\,\cc\ with 
the largest distance from the dark matter halo center. 
One can indeed see a significant offset between the gas and dark matter 
centers. However, the high-density gas region is still within 
both the half-mass as well as the virial radius of the dark matter halo. 

We identify high density gas regions by searching the gas in the simulation 
box for gas cells that exceed 200 times the critical density of the 
Universe.\footnote{The resulting number density thresholds in neutral atomic gas are $n_\mathrm{thres} = 1.1 $\,cm$^{-3}$ at redshift $z=15$ and $n_\mathrm{thres} = 2.4$\,cm$^{-3}$ at redshift $z=20$.} 
We make a list of these cells and start to work our way through it from highest to lowest density. Each cell on the list is taken to be the center of a separate high-density region with a radius of 100 pc (roughly the virial radius of a halo of $M=10^5$\,M$_\odot$ at these redshifts). To minimize double-counting, we remove from the list any cells that lie within an already-identified high-density region, which guarantees that the centers of these regions have a separation of at least 100~pc.\footnote{Note that there may still be some overlap if the centers of two regions are separated by less than 200~pc. In this case, the gas in the overlapping region is attributed to each region.} In a final step, we remove gas clumps that have a higher average density than their central density, because these simply trace the outskirts of a neighboring regions with a high density. In practice, this only happens for gas clumps with central densities close to the density threshold, $n_\mathrm{thres}$. None of the gas regions with $n_\mathrm{max} \ge n_\mathrm{SF} = 100\,\mathrm{cm}^{-3}$ are affected. 

At the end of this process, we have a list of independent high-density regions, located at various distances from their closest associated dark matter haloes. In the following section, we will use the terms ``high-density region'' and ``gas clump'' to refer to regions in this list and the term ``proto-globular cluster'' to refer to the subset of these regions that lie outside of the half-mass radius of their nearest halo.
\section{Analysis}
\label{sec:stat}
\subsection{Distance between high-density gas regions and halos}
\begin{figure*}
\centering
\includegraphics[width=.98\columnwidth]{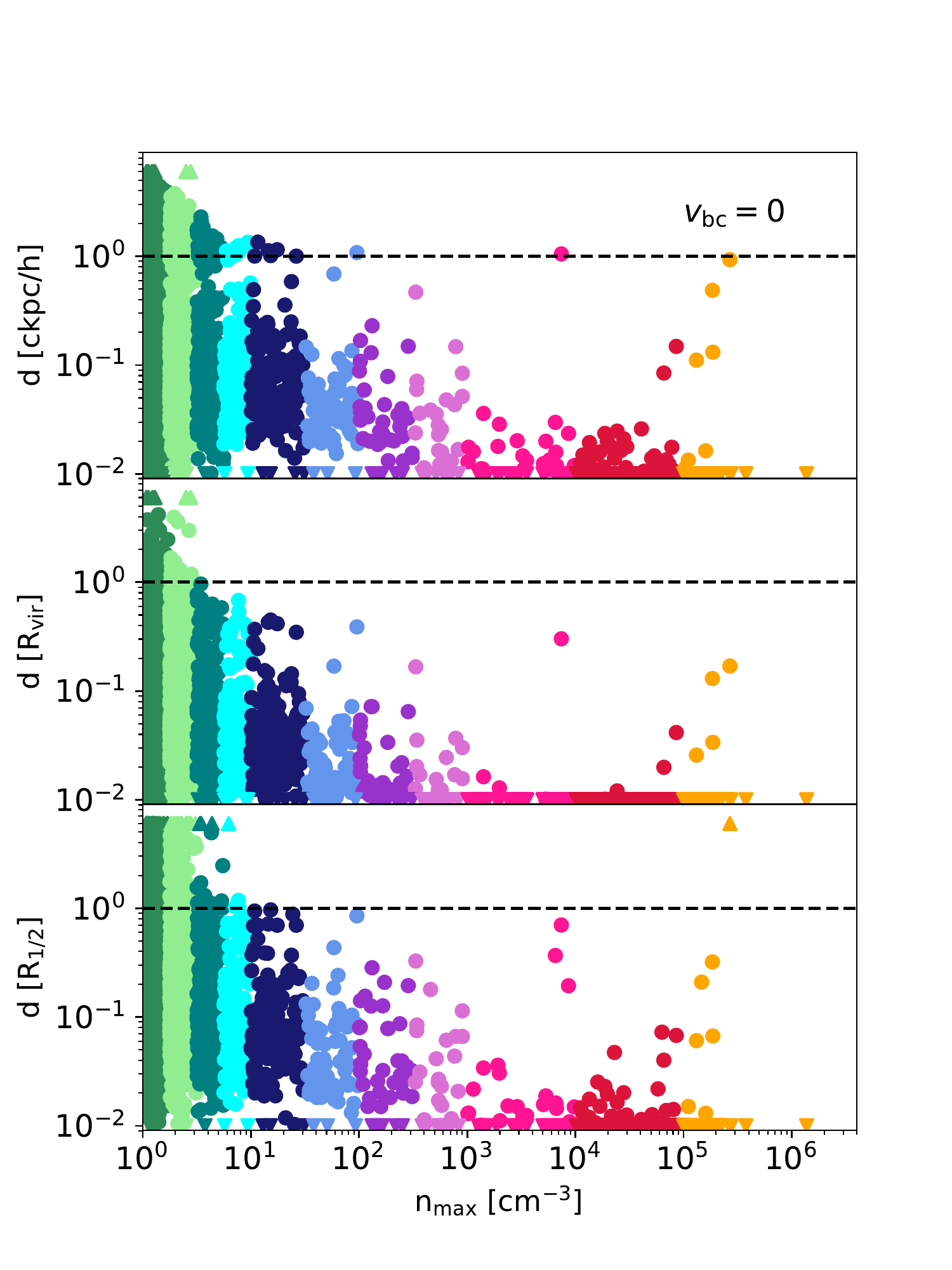}
\includegraphics[width=.98\columnwidth]{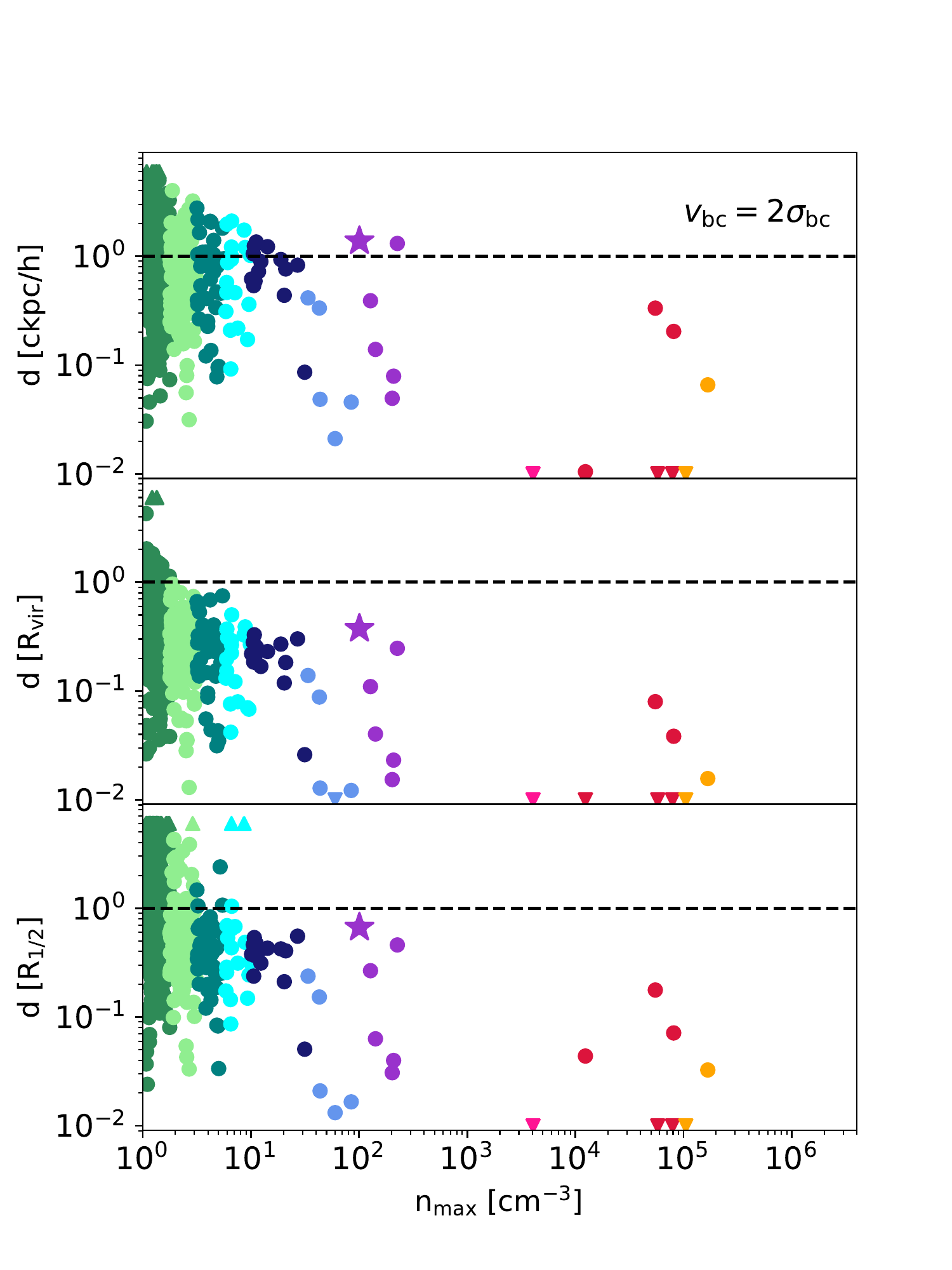}
\caption{Distance between the high-density gas region and the closest halo as a function of the maximum gas density 
at redshift $z=15$. Left panel: no streaming velocity; right panel: streaming velocity of 2$\,\sigma_{\rm bc}$. In the top row, we show the distance in ckpc$/h$, in the middle row as a fraction of the virial radius and in the bottom panel as a fraction of the half-mass radius. The data points are color-coded by density, from lowest density (green) to highest  density (orange). The region highlighted in Figure \ref{fig:slice} is marked with a purple star in the right-hand panels. }
\label{fig:dist_ndens}
\end{figure*}

\begin{figure}
\centering
\includegraphics[width=.98\columnwidth]{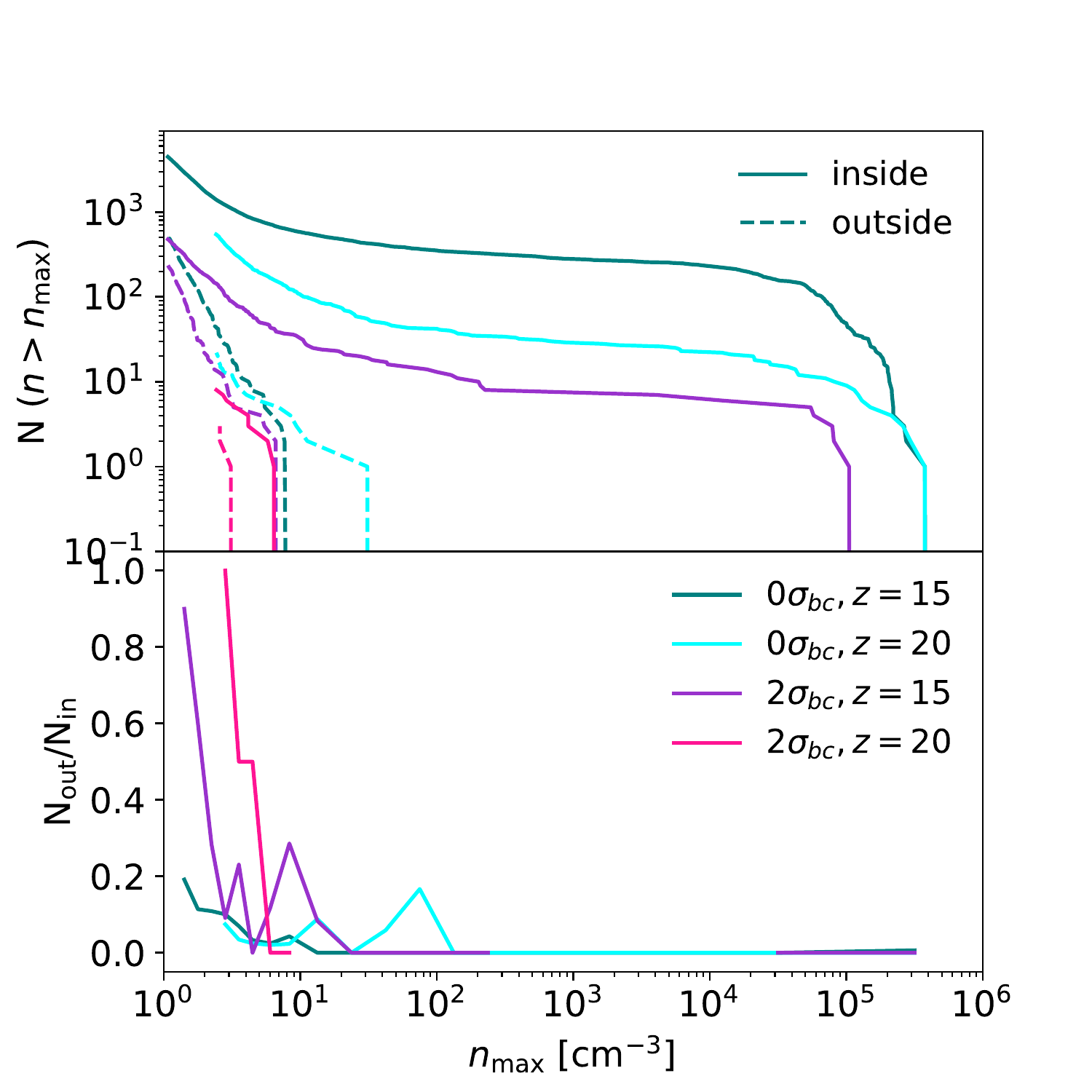}
\caption{Top panel: cumulative number of high density regions inside (solid lines) or outside the half mass radius (dashed lines) of a dark matter halo. 
Bottom panel: fraction of high-density regions inside  / outside the 
half-mass radius of the halos. }
\label{fig:number}
\end{figure}

\begin{figure*}
\centering
\includegraphics[width=.98\columnwidth]{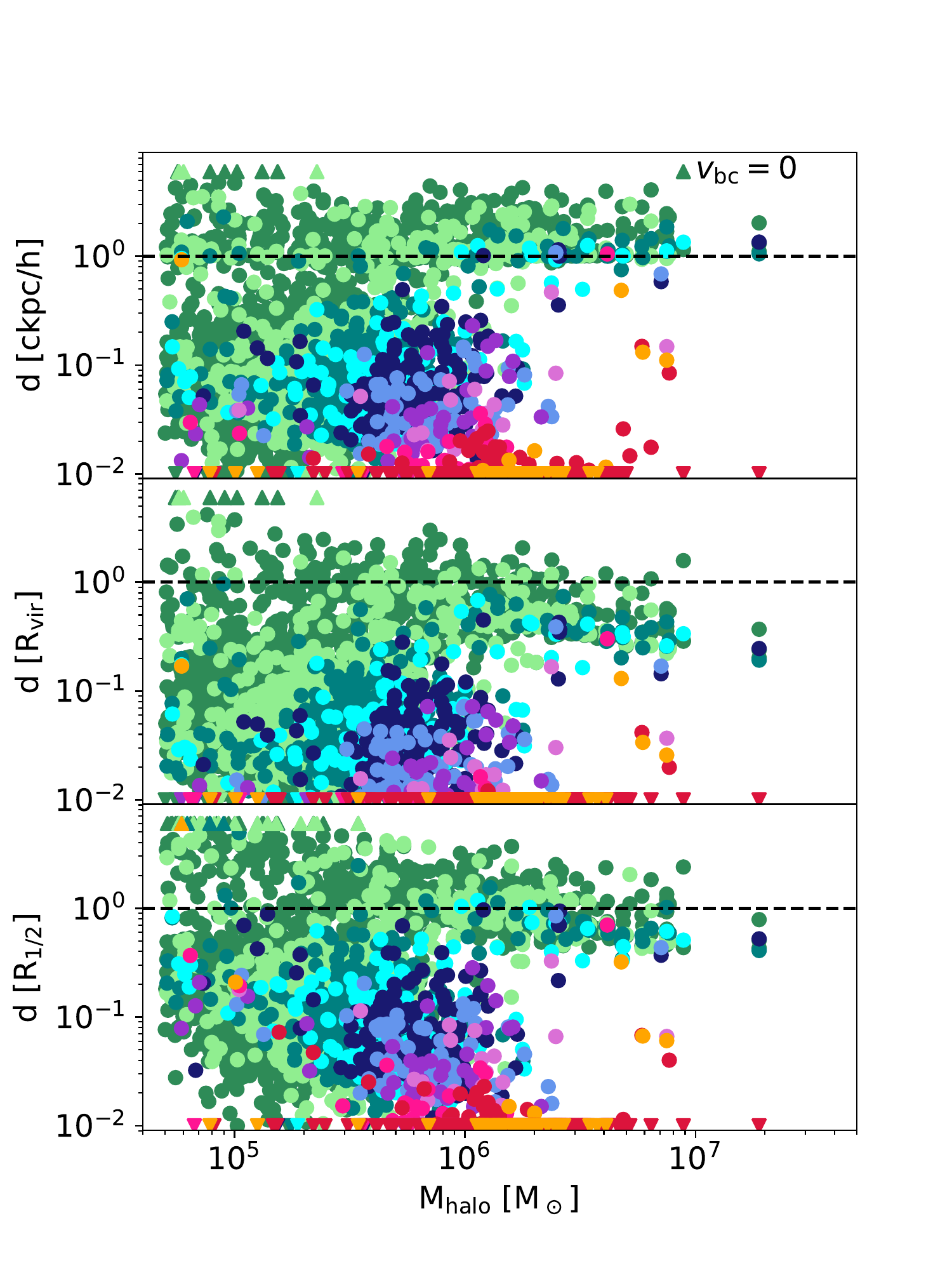}
\includegraphics[width=.98\columnwidth]{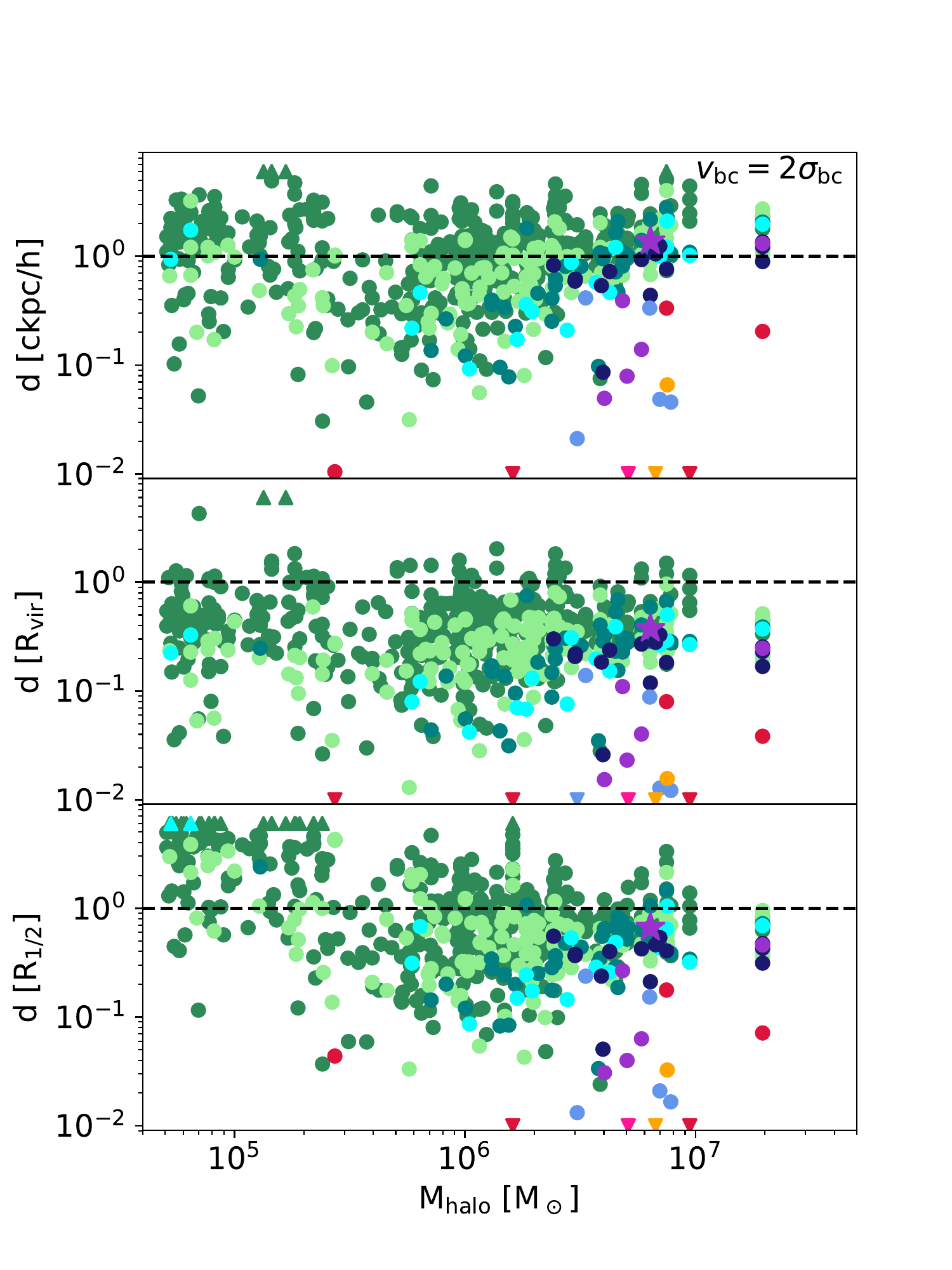}
\caption{Distance between the high-density gas region and the closest halo as a function of the halo mass at redshift $z=15$. Left panel: no streaming velocity, right panel: streaming velocity of 2$\sigma_{bc}$. In the top row, we show the distance in ckpc$/h$, in the middle row as a fraction of the virial radius  and in the bottom panel as a fraction of the half-mass radius. The data points are color-coded by density, from lowest density (green) to highest  density (orange). }
\label{fig:dist_mass}
\end{figure*}

We begin our analysis by investigating whether the high-density gas regions 
lie within the virial or half-mass radius of the halo closest to them.
In Figure \ref{fig:dist_ndens}, we show the distance between each high-density region and the halo closest to it in multiples of the virial- and the half-mass radius at redshift $z=15$. 
The simulation without streaming  velocities (left) has a significantly 
higher number of high-density regions than the simulation with a 
streaming velocity of 2\,\s. 
This is not surprising, as the number of star-forming halos decreases 
for increasing streaming velocities \citep{schauer20}. 

To identify gas that is likely to form stars in the near future, we follow \cite{schauer19} and adopt a density threshold of $n_\mathrm{SF} = 100$\,cm${^{-3}}$. This density is several hundred times larger than the mean baryon density in a virialized halo and hence is only likely to be reached in gas which is cooling and undergoing gravitational collapse. Looking at Figure \ref{fig:dist_ndens}, we see that none of the high density regions with $n > n_\mathrm{SF}$ (denoted by the points with purple, red or orange colors) lie outside either the virial or the half-mass radius of their nearest halo.
Only gas regions with a number density of 10\,cm${^{-3}}$ (3\,cm${^{-3}}$) or lower 
are found outside of the half-mass (virial) radius. 
This finding is independent of the presence of a streaming velocity, 
as can be seen by comparing the left and right panels of Figure 
\ref{fig:dist_ndens}. 
The best candidate for a future globular cluster -- the region with $n > n_{\rm SF}$ that is furthest from its associated halo -- is shown as a purple star in the right panels of Figure \ref{fig:dist_ndens}. This candidate region lies just within the half-mass radius of its associated halo, but this places it well within the virial radius of the halo (see also Figure \ref{fig:slice}).

The same is true if we look at the simulations at an earlier time, $z=20$. At this redshift, no gas with a number density exceeding 10\,cm$^{-3}$ is found anywhere in the streaming velocity simulation, and the highest gas number densities reached outside of the half-mass radius of any halo are a few cm$^{-3}$. In the non-streaming 
velocity simulation, gas reaches high densities, but again only the 
lower density gas regions lie outside the half-mass ($n < 100$\,cm$^{-3}$) or virial radius ($n < 10$\,cm$^{-3}$) of the 
corresponding dark matter halos. 

High-density regions are more likely to lie away from the halo center 
for a high streaming velocity than for no streaming velocity. 
This can be seen more quantitatively in Figure \ref{fig:number}. 
Here, we show the cumulative number of high-density regions as a function of the maximum density of the region that are located inside (solid lines) or outside (dashed lines) the half-mass radius of the closest dark matter halo.

As expected, the total number of halos increases as the redshift decreases. We also see a clear impact of the streaming velocity: higher streaming velocities lead to the formation of a smaller number of halos at a given redshift, as already noted in several previous studies \citep{naoz12,popa16,schauer19}. 
As already indicated, we find no regions with $n > 100 \: {\rm cm^{-3}}$ outside of the half-mass radius of its associated halo, regardless of whether or not a streaming velocity is present. If we look at the dense clumps with $n < 100 \: {\rm cm^{-3}}$, however, we see a clear difference in behavior between the simulations with and without streaming. In the run without streaming, most of these clumps are located within the half-mass radius, with only $\sim$20\% being located outside of this radius. On the other hand, in the run with  2\,\s\ streaming, more than half of these clumps are found outside of the half-mass radius of their nearest halo.

In Figure \ref{fig:dist_mass}, we show the distance between the densest gas peaks and the 
dark matter halos as a function of the halo mass. The color coding is the same as in 
Figure~\ref{fig:dist_ndens}, with points being colored by 
density from dark green to orange. Colors with a red component (purple, pink, red, orange) are used to denote regions with densities $n > n_{\rm SF}$.
One can see that there is little correlation between the distance and the halo mass, 
and that high densities are primarily found in higher mass halos. 
Note that in comparison to \cite{schauer20}, we use a different halo mass definition 
and therefore the masses here for the high-density gas regions can lie below the 
masses determined as minimum gas mass for star formation in a halo. 
One can also see from the data points representing the highest halo masses that 
some halos host several high-density regions. 

\subsection{Mass in high-density regions}
\begin{figure}
\centering
\includegraphics[width=.98\columnwidth]{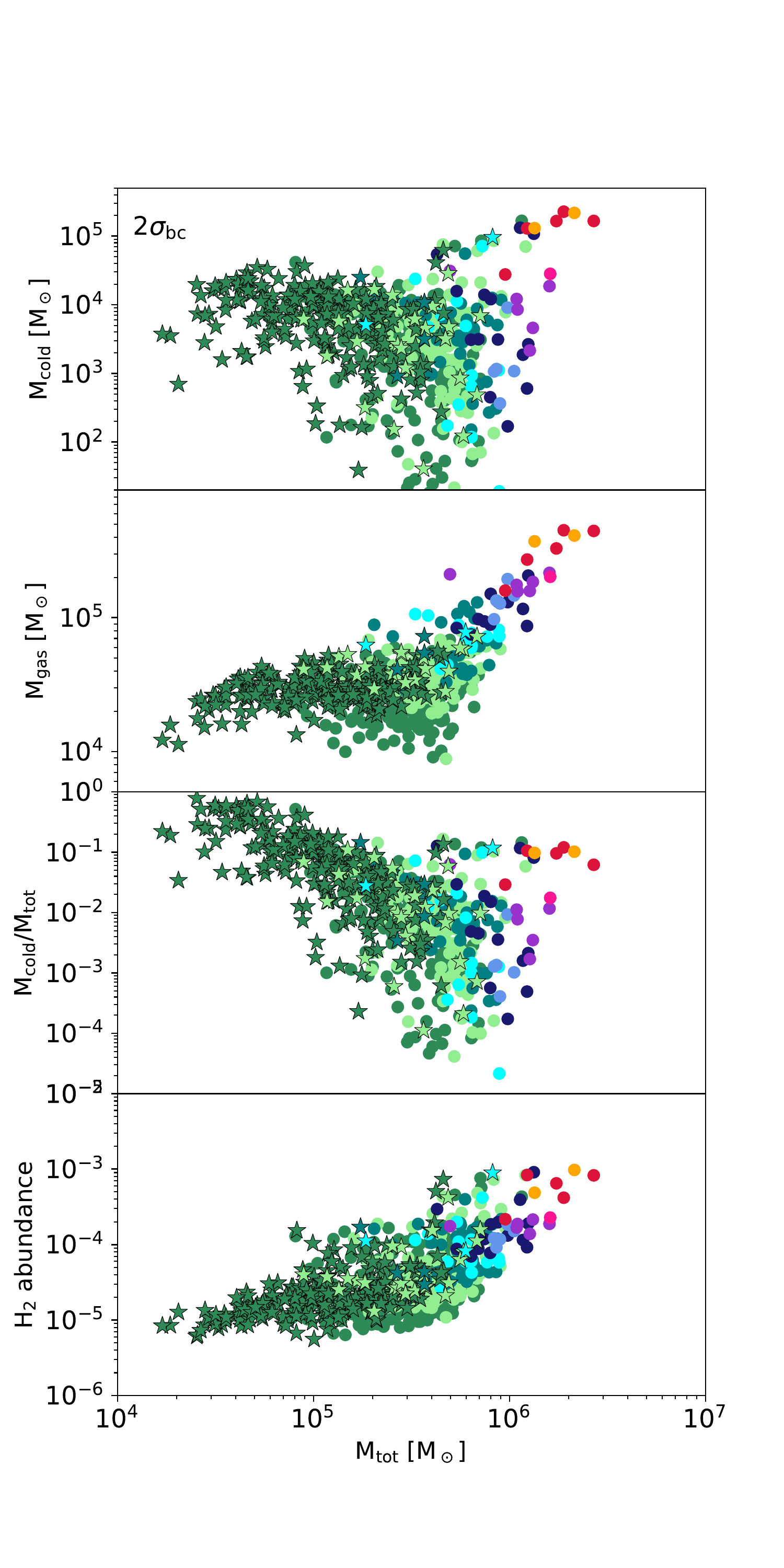}

\caption{Properties of the gas in high-density regions as a function of the total (gas and dark matter) mass in this region.
Stars show regions which lie outside the half-mass radius of the closest halo. The color coding refers to the maximum density of the gas region, and is the same as in Fig.~\ref{fig:dist_ndens}, from dark green (lowest gas density) to orange (highest gas density). From top to bottom, the panels show the cold ($T < 500$~K) gas mass, the total gas mass, the cold gas fraction ($M_{\rm cold} / M_{\rm gas}$) and the mean H$_{2}$ abundance.
One can see that gas regions with a very high cold gas fraction often 
lie outside the half-mass radius of the closest dark matter halo, 
but the largest cold gas mass can be 
found in the most massive regions, which often have the highest densities. 
The large cold gas masses seen in the top panel are caused by cooling by molecular hydrogen, as the gas in high-mass halos has 
a large H$_2$ abundance (bottom panel). }
\label{fig:masses}
\end{figure}

In order to determine whether any of the high-density regions selected from our simulations could potentially be the progenitors of globular clusters, we need to look not only at the maximum density reached in the region, but also its mass and ratio of gas to dark matter. We have therefore calculated the gas mass, total mass (gas and dark matter combined) and mass of cold ($T \le 500$\,K) gas associated with each high-density clump. 

In Figure \ref{fig:masses}, we show the cold gas mass ($M_{\rm cold})$, the total gas mass ($M_{\rm gas}$) and the cold gas fraction ($M_{\rm cold} / M_{\rm gas}$) as a function of the total mass ($M_{\rm tot}$) for all of the high-density regions identified in the 2\s\ streaming velocity simulation at redshift $z=15$. Following \cite{chiou18}, objects that lie outside the half-mass radius of the closest halo -- the most plausible candidates to be proto-globular clusters -- 
are indicated by stars, while all other gas regions are indicated by
dots. We adopt the color coding according to the maximum density from 
Figure \ref{fig:dist_ndens}, where reddish colors show a number density 
of 100\,\cc\ or higher. 

The gas masses of the high-density regions range from $10^{4} \: {\rm M_{\odot}}$ up to $\sim 5 \times 10^{5} \: {\rm M_{\odot}}$. However, the regions with the highest gas masses are all located within the half-mass radius of their associated dark matter halo; the most massive regions outside of the half-mass radius have $M_{\rm gas} \sim 10^{5} \: {\rm M_{\odot}}$. 

The cold gas mass covers a much broader range of values, from a few tens of \Ms\ up to $\sim 10^{5} \: {\rm M_{\odot}}$. It is highest in regions with a large total mass, 
$M_{\rm tot} \ge 10^6$\,\Ms. These regions host the highest density 
gas, which can be seen by the reddish colors of the points 
in the upper right corner of the top panel. 
Cooling by molecular hydrogen is effective in these halos, and reduces the temperature of the central high-density gas to below 500\,K. 
One can see that in the high-mass halos with associated high density gas, the abundance 
of molecular hydrogen is indeed much higher than in the low-mass halos 
(bottom panel). 
This is reflected in a relatively high cold gas fraction (2nd panel from the bottom), 
ranging from a few percent to more than 10\,\%. 
As in \cite{chiou20}, the cold gas fraction is highest in objects with a gas fraction $\ge$ 40\% (shown with stars). However, the cold gas in these objects is not a product of H$_2$ cooling. Rather, it is a consequence of the low mass associated with these regions. The virial temperature in a halo with $M_{\rm tot} < 10^{5} \: {\rm M_{\odot}}$ is less than 500~K, so the gas in these regions is cold by our definition because it has never been heated up above 500~K, rather than because it has cooled down from a higher temperature.

\begin{figure*}
\centering
\includegraphics[width=1.98\columnwidth]{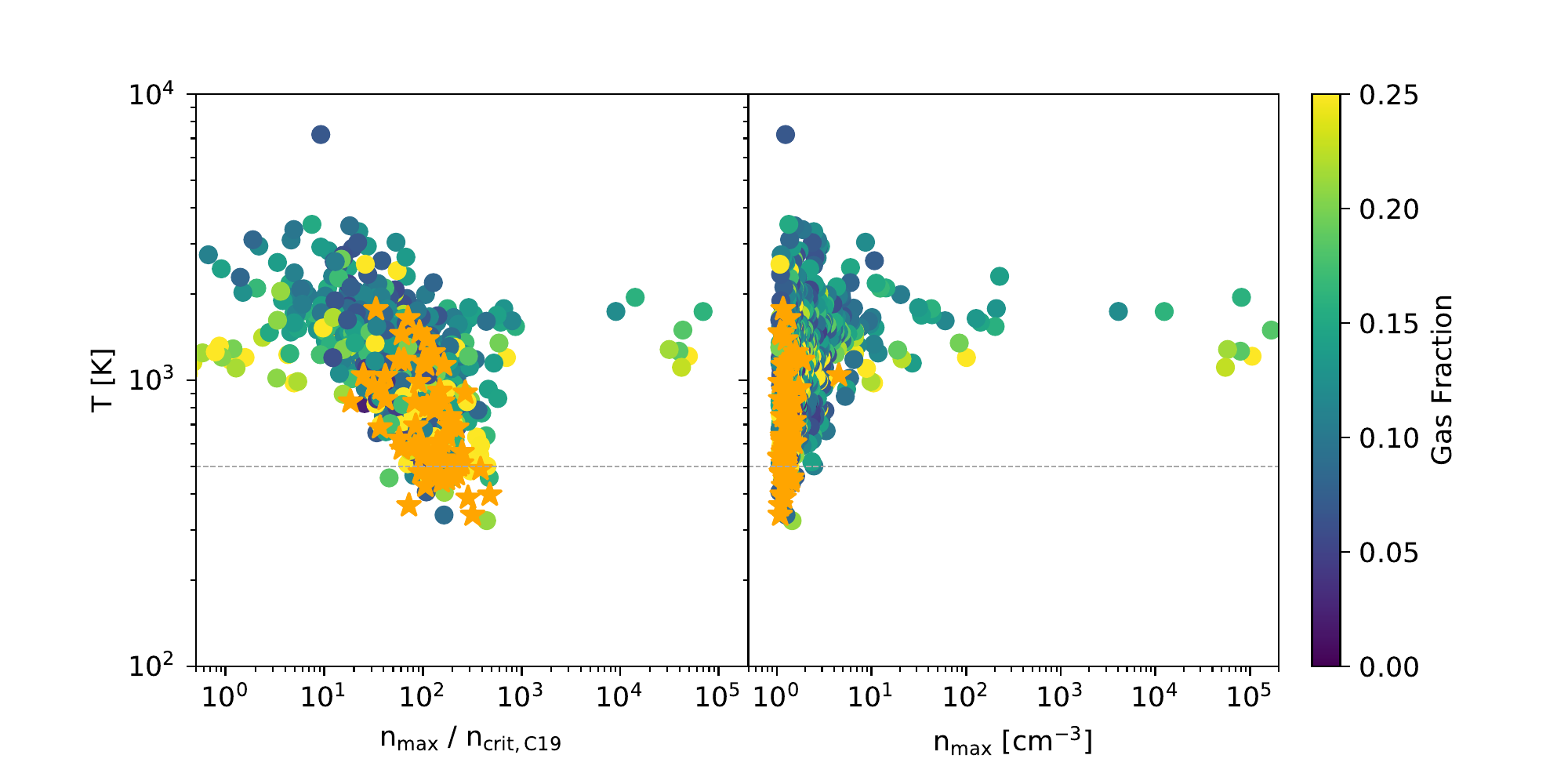}
\caption{Average temperature as a function 
of the number density for the gas clumps. 
We mark the $T = 500$~K threshold in Fig.~\ref{fig:masses} with a dotted line. 
The points are color coded by gas fraction, with orange stars 
displaying gas regions with 
a gas fraction of 40\% and higher. One can see that while these 
high gas fraction regions are super-critical according to 
the definition of \cite{chiou19}, they correspond to low maximum 
density regions. }
\label{fig:temp}
\end{figure*}

The mass-weighted average temperature of the high-density gas regions can also be 
seen in Figure \ref{fig:temp}, where we show the temperature as a 
function of the maximum number density. This time, the color coding 
represents the gas fraction in the dense region, and objects with 
a gas fraction $\ge$\,40\% are shown by orange stars. 
One can see that these objects are all associated with a relatively 
low maximum density (right panel), while the gas fraction of regions with a high maximum density varies from a few percent to slightly above the cosmic mean. 

Although we find a broad range of different temperatures in regions with densities close to $n_{\rm thres}$, indicative of the wide variety of different dynamical histories of this gas, above $n \sim 10 \: {\rm cm^{-3}}$ all of the average temperatures lie in the range 1000--2000~K, with no dependence on density. This uniformity of temperature indicates that this gas is being cooled efficiently by emission from H$_{2}$: if it were evolving adiabatically, then we would expect an increase in temperature by a factor of 100 between $n = 100 \: {\rm cm^{-3}}$ and $10^{5} \: {\rm cm^{-3}}$ as opposed to the flat distribution of values we actually find. 
\subsection{Jeans mass and collapse criteria}
A necessary condition for the high-density regions that we have identified to be able to gravitationally collapse is that they not be supported by the supersonic turbulent motions in the gas. \citet{chiou19} formulate this condition in terms of the Jeans length $\lambda_{\rm J} = (\pi\,c_s^2 / G \rho)^{1/2}$ and the sonic scale of the turbulent flow,  $\lambda_{\rm s} = L_\mathrm{drive}/{\cal M}^2$, where $L_{\rm drive}$ is the driving scale of the turbulence and ${\cal M}$ is the Mach number of the turbulence on this scale. \citet{chiou19} argue that collapse is only possible if $\lambda_{\rm J} \leq \lambda_{\rm s}$, or, equivalently, if the density of the high density region exceeds a critical value 
\begin{equation}
\rho_\mathrm{crit, C19} = \pi \, \frac{c_s^2 \, {\cal M}^4 } {G\,L_\mathrm{drive}^2}\,.
\end{equation}
To calculate this value and the corresponding critical number density $n_\mathrm{crit, C19}$, we first calculate the sound speed of the high-density region by averaging over the local sound speed of each gas cell belonging to the region. We also
calculate the 3D velocity dispersion for the same region. The
Mach number ${\cal M}$ then follows as the ratio of these two quantities. We take the driving scale of the turbulence to be equal the diameter of the high density region. 

In the left panel of Figure \ref{fig:temp}, we plot the temperature of each of our high-density regions as a function of their maximum density divided by this critical density $n_\mathrm{crit, C19}$. One can clearly see that almost all gas regions exceed this critical density threshold, i.e.\ we do not expect turbulent motions to impede their collapse. For low and high values of $n_\mathrm{max} / n_\mathrm{crit, C19}$, the mean temperature is around 1000~K, but for intermediate values of $n_\mathrm{max} / n_\mathrm{crit, C19}$ close to 100, there is a much broader distribution of temperatures, with some regions having mean temperatures below 500~K (denoted by the thin grey line). In particular, a number of regions with high gas fractions (40\,\% or higher, denoted as orange stars in the figure) have these low temperatures. 

However, we can see that these are low maximum density regions (right panel) in halos with low virial masses, which are likely cool not because of H$_2$ cooling but because they have never been heated to high temperatures.
A direct comparison between our work and \cite{chiou19} is not straightforward, as we are not using their ellipsoid fitting procedure for our high density regions. However, we have verified that our result is robust to the size we choose for our high density regions (i.e.\ we do not find qualitatively different results if we increase or decrease the adopted radius from our default value of 100~pc). In each case, the cold gas mass is largest in gas clumps that are located inside the half-mass radius of the halo, and these are also the only regions that have number densities $n > n_{\rm SF}$ and that show evidence for H$_{2}$ cooling. The dense regions located outside of the half-mass radius do not appear to be collapsing or cooling efficiently even though they exceed the critical density $n_\mathrm{crit,C19}$. 
\begin{figure}
\centering
\includegraphics[width=1.12\columnwidth]{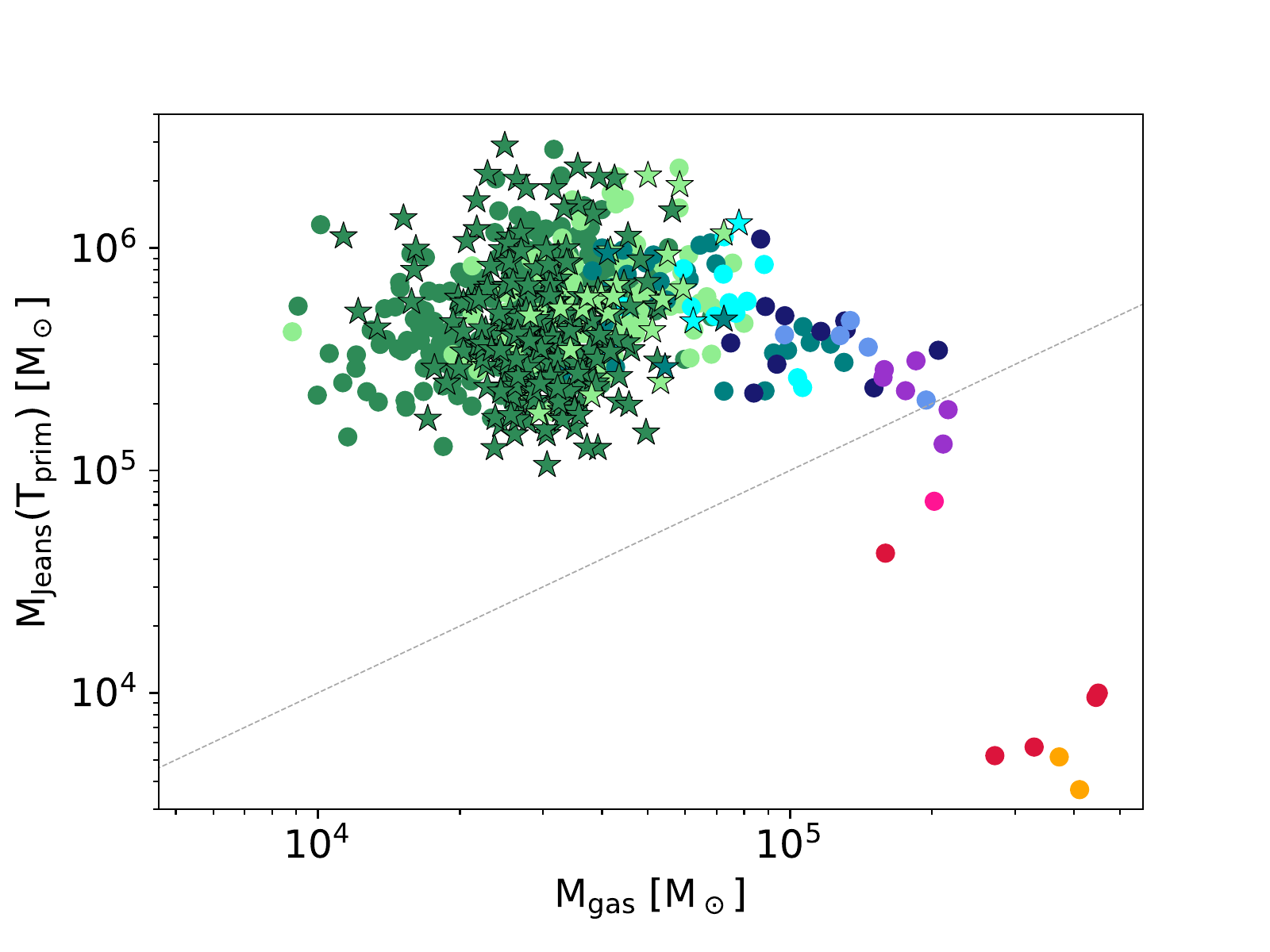}
\caption{Jeans mass as a function of the gas mass in a high-density region. Only the highest density regions (shown with a reddish color) have Jeans masses below the gas mass, the gray dashed line shows the 1:1 relation. No gas clump that lies outside the half-mass radius of its dark matter halo is Jeans-unstable.}
\label{fig:mgas-mjeans}
\end{figure}

To further determine whether the dense regions we find outside of the half-mass radius of the dark matter halo are able to collapse, we compare the gas mass of the high-density clump to the Jeans mass  
\begin{equation}
M_J = \frac{\pi}{6} \frac{c_s^3}{G^{3/2}\rho^{1/2}}. 
\end{equation}
Here, $c_s$ is the sound speed, $G$ the gravitational constant, and 
$\rho$ the density. 

In Figure \ref{fig:mgas-mjeans}, we show the Jeans mass as a function of the gas mass 
of all gas regions in our 2\,\s\ simulation at $z=15$. 
One can clearly see that only the gas regions with the highest densities 
(denoted in colors that have a red component, with orange being the densest) 
have a gas mass that exceeds the Jeans mass for their density and temperature. 
No region outside of the half-mass radius of their dark matter halo (shown by stars) is found below this threshold. 

\begin{figure*}
\centering
\includegraphics[width=1.98\columnwidth]{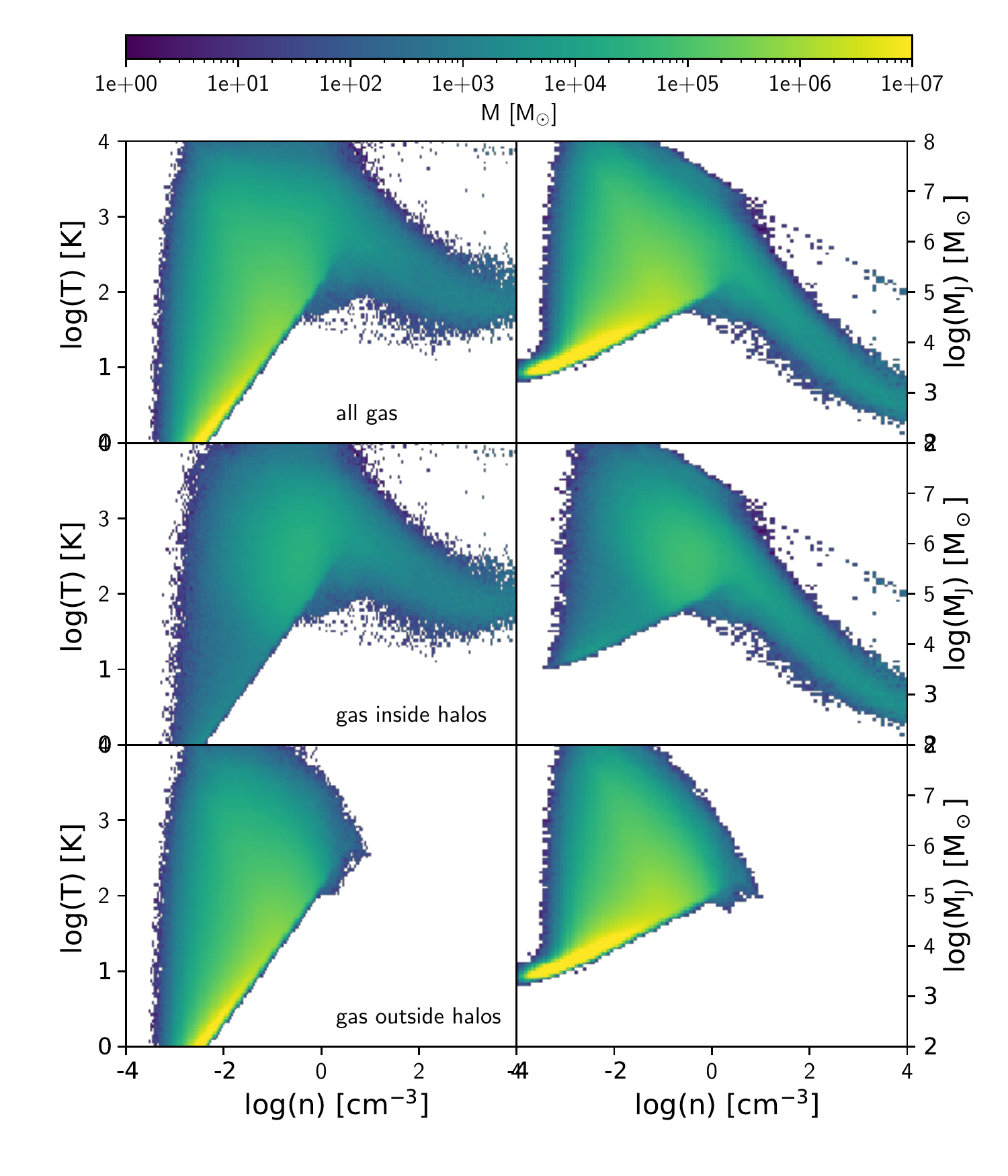}
\caption{Temperature (left) and local Jeans mass (right) of the full simulation box v2\_lw0 at redshift $z=15$, shown as a function of density. In the top row, we show all gas;
in the middle row, we only show gas that is within the half-mass radius of any halo; and in the bottom row, we show the gas that is outside the half-mass radius of all halos. 
The phase-space diagram for all gas shows a typical distribution for 
metal-free gas, with H$_2$ cooling setting in at a number density of $n \approx 1$\,cm$^{-3}$, with temperatures of a few hundred K at high  number densities. 
Most of the dense gas is inside of the halos. 
However, some of the high-density gas ($n >1$\,cm$^{-3}$) lies outside the 
half-mass radii of the halos, reaching Jeans masses of 10$^5$\,\Ms\, and below.}
\label{fig:hist}
\end{figure*}

In addition to determining the Jeans mass for each gas region individually, 
we also analyze the local Jeans mass in the whole simulation box. 
We show the phase-space diagram of the gas region in the left column of Figure \ref{fig:hist}, 
and the local Jeans mass -- calculated from the density and temperature of the respective 
gas cell -- in the right column. 
We divide the gas in the box into gas within a dark matter host halo (within the half-mass radius, middle row) and gas outside of the dark matter halos (bottom row). 
One can see that the gas in the simulation follows the typical high-redshift 
cooling curve \citep[see e.g.][]{yoshida06}, which 
shows the onset of H$_2$ cooling at $n\approx1$\,\cc, after 
which the temperature drops to below 500\,K for the higher density 
gas. This high-density gas is almost entirely found inside 
of dark matter halos, as no high density gas outside of dark matter halos can be seen in the bottom left panel of Figure~\ref{fig:hist}. 

Gas can collapse and ultimately form stars when the local Jeans mass is small. The high-density gas, cooled by molecular hydrogen,
has a low local Jeans mass, below 100\,\Ms. However, 
the local Jeans mass drops to these small values only for 
gas that resides inside of dark matter halos, as can be seen 
in the middle right panel of Figure \ref{fig:hist}. 
High density gas outside of dark matter halos still has a very high local Jeans mass, exceeding $M_\mathrm{J} \ge 10^4$\,\Ms. 
\textit{Globally, we find that primordial gas is only able to cool and form stars when inside a dark matter halo. }
\subsection{Metal-enriched cooling}
\label{sec:metals}
In the previous section, we saw that although a high streaming velocity leads to the formation of many high-density clumps outside of the half-mass radius of any halo, these clumps have masses lower than the Jeans mass and hence are gravitationally stable. In order for these clumps to become gravitationally unstable, we would have to decrease their Jeans masses by 1--2 orders of magnitude. At fixed density, this corresponds to a decrease in their temperature by approximately an order of magnitude, meaning that the gas would have to reach temperatures in the range of 50--100~K. 

Reaching these temperatures with H$_{2}$ cooling alone is impossible. Even in optimal conditions, H$_{2}$ cooling struggles to reduce the temperature much below 200~K, owing to the exponential decrease in the cooling rate that occurs below 512~K. Moreover, the conditions in these clumps are far from optimal for efficient H$_{2}$ cooling: as we have already seen, they have relatively low H$_2$ abundances and densities far below the H$_{2}$ critical density. However, we know that in reality, globular clusters are not completely metal-free: they have low, but non-zero,  metallicities \citep{harris96}. It is therefore worthwhile to ask what level of metal enrichment would be required in order for these regions to cool sufficiently to become gravitationally unstable, and whether this is consistent with what we know about globular cluster metallicities.

Since we are interested in developing an order of magnitude estimate of the required metallicity, we simplify the question by asking what level of metal enrichment is required to cool the gas to the CMB temperature, which varies from $\sim 40$ to 60~K at the range of redshifts considered here. The actual metallicity required to cool any given high density clump sufficiently that $M < M_{\rm J}$ will differ somewhat from this estimate, but not by a large amount. 

Our starting point is the Rees-Ostriker criterion \citep{rees77}, which states that a gas cloud is able
to cool and collapse if its cooling time is smaller than its free-fall time. 
Equating the free-fall time $t_\mathrm{ff} = 1/\sqrt{G \rho}$ with 
the cooling time 
$t_\mathrm{cool} = n k_{\rm B} T / [(\gamma -1) \Lambda]$
leads to a minimum required cooling rate of 
\begin{equation}
\Lambda_\mathrm{req} = \frac{3 k_{\rm B}}{2} \sqrt{G \mu m_{\rm H}} T n^{3/2}, 
\end{equation}
where $k_{\rm B}$ is the Boltzmann constant, 
$\mu$ is the mean molecular weight, $m_{\rm H}$ is the mass of a hydrogen atom, and $\gamma$ is the adiabatic index. Since we expect the primordial gas to be predominantly neutral and atomic, we have $\gamma = 5/3$ and $\mu = 1.22$. 

\begin{figure}
\centering
\includegraphics[width=.98\columnwidth]{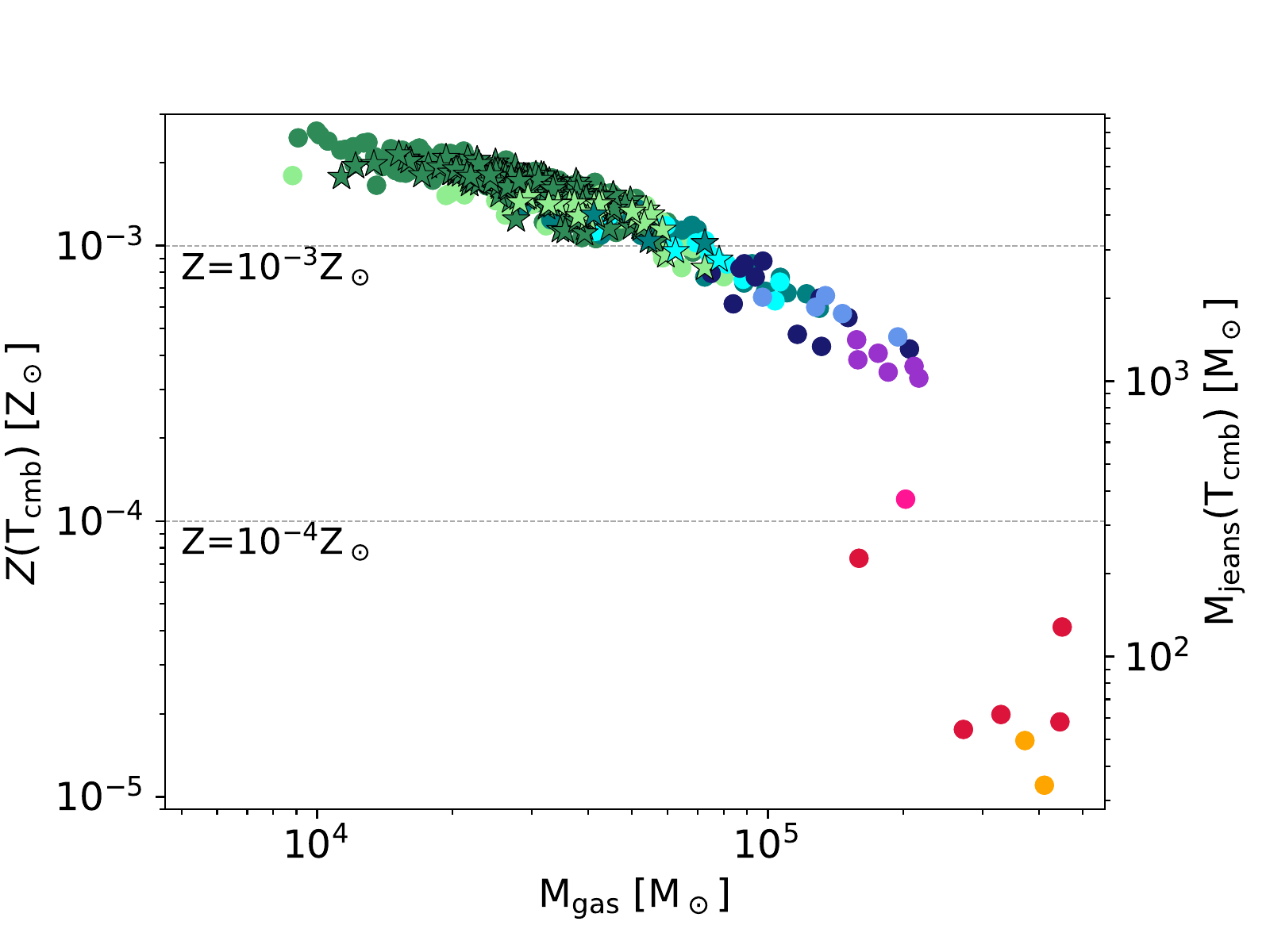}
\caption{Required metallicity (left $y$-axis) for each gas region to cool to the CMB temperature as a function of gas mass. Most gas regions require a metallicity of about $10^{-3}\,Z_\odot$, although the high density regions (reddish) colors that are Jeans unstable with primordial cooling alone require lower values. 
In the same figure, we show the Jeans mass that these regions would have if they had a temperature $T = T_{\rm CMB}$ (right $y$-axis). All of our identified high density regions can become Jeans unstable under this assumption, as they are below the 1:1 correspondence line (gray dashed). Gas regions outside the half-mass radius are shown with stars, gas regions within the half-mass radius of the closest halo are shown with circles.}
\label{fig:cmb}
\end{figure}

The cooling rate at high redshift depends on the temperature, density and 
metallicity, $Z$, of the gas. We use the approximation in \cite{bromm01b} as a convenient analytical fit to $\Lambda$, roughly valid over the parameter range of interest here: 
\begin{multline}
\Lambda (Z) \approx 3\times10^{-30} \mathrm{erg\,cm^{-3}\,s^{-1}} \times    \\
\left(\frac{Z}{10^{-4}\: \mathrm{Z}_\odot} \right)
\left(\frac{n}{1 \: \mathrm{cm}^{-3}} \right)^{2}
\left(\frac{T}{5000\,\mathrm{K}} \right)^{1/2}. 
\end{multline}
Combining these expressions, we can estimate the required metallicity 
as a function of gas temperature and density:
\begin{equation}
Z_{\rm req} = 1.80\times10^{-4} Z_\odot 
\left(\frac{n}{\mathrm{cm}^{-3}} \right)^{-1/2}
\left(\frac{T}{5000\,\mathrm{K}} \right)^{1/2}. 
\label{equ:z}
\end{equation}
We know from Figure \ref{fig:mgas-mjeans} that only the high-density regions (shown in orange to purple colors) are Jeans unstable. 
However, if metal cooling as described above is successful and 
can lower the temperature in a gas region to the CMB floor, $T = T_\mathrm{CMB}(z) = 2.73\,(1+z)\,\mathrm{K}$, 
the Jeans mass is reduced correspondingly. 
In Figure~\ref{fig:cmb} (right vertical axis), we show that 
all our gas regions have a Jeans mass below the gas mass, and are hence Jeans unstable if cooled down to the CMB temperature. 

The metallicity to cool the gas regions to such a low temperature -- as determined 
in Equation~\ref{equ:z} -- is also presented in Figure~\ref{fig:cmb} (left vertical axis).
We find that metallicities of about $Z\approx10^{-3}Z_\odot$ are required for cooling 
proto-globular clusters down to the CMB floor, allowing them to collapse. 
Metal enrichment from Pop~III stars can occur either through core-collapse or 
pair-instability supernovae, depending on the mass of the Pop~III progenitor. 
\cite{jeon14b} and \cite{chiaki19} find that core-collapse supernovae with progenitor masses of $\sim10-40$\,\Ms\ may enrich a $10^5 - 10^6$\,\Ms\,host halo to metallicities of a few $10^{-4}$\,Z$_\odot$, with a re-collapse time of $\sim10-100$\,Myr. 
\cite{ritter12}, on the other hand, find that a minihalo with an initial mass of $10^6$\,\Ms\ in which a 40\,\Ms\ star explodes re-accretes gas with a metallicity of 0.01--0.001\,$Z_\odot$ after only a few tens of Myr. 

Pair-instability supernovae (PISNe) have progenitor masses between 140 and 
260\,\Ms, and explode with a factor 10-100 times higher explosion energy 
of $10^{52-53}$\,erg, releasing up to 100\,\Ms\,of metals into their
surroundings \citep{hw02}. 
Consequently, the metallicity of the host halos of these supernovae is larger. 
For example, \cite{wise08} find that PISNe enrich their host halos to metallicities 
$Z \approx 10^{-3}Z_\odot$, with supernova remnants and the low density intergalactic medium containing the most metals. 
In a simulation by \cite{greif10}, a pair-instability supernova does not only 
enrich the host galaxy to metallicities of $Z > 10^{-3}\,Z_\odot$, 
but also two neighboring minihalos to metallicities of $Z > 10^{-3.5}\,Z_\odot$. 
As these simulations are computationally expensive, typically only one or a few 
halos are studied. The three dimensional structure of minihalos can vary substantially  \citep{yoshida03,druschke18}, and it is not surprising that the enriched halos 
vary in metallicity. 

We conjecture that, in the context of globular cluster formation, external enrichment would play the dominant role in enriching dense regions far from the halo center with metals, since we are trying to determine the critical metallicity for a gas region to collapse for the first time. 
The required levels of metallicity found in our study are readily provided by the metal enrichment from pair instability supernovae, as well as from core-collapse events \citep{ritter12}. The requirement is more easily met by internal rather than external enrichment, 
as supernova explosions generally enrich neighboring halos to lower values. 
However, none of the above studies include streaming velocities, which might 
enable supernovae explosions to distribute their metals further into 
the intergalactic medium since the gas overdensities are more 
rare in streaming velocity regions. More work on the simulation side, including the combined effects of streaming velocities, primordial chemistry, and low-metallcity cooling, is required to understand 
whether globular clusters can form through this  channel. 
\section{Observational Signature}
\label{sec:dis}
In this work, we have investigated whether streaming velocities produce high-density gas regions 
outside of dark matter halos that could be the progenitors of globular clusters, 
as has been suggested by \cite{naoz14}. 
This could be a natural explanation for the old age of globular clusters, linking them to the epoch of first star formation, and for their lack of a dark matter component. We will briefly explore possible observational consequences. 

As streaming velocities are distributed in the Universe over scales of several Mpc, 
we would expect this distribution to be reflected in 
metal-poor globular clusters if this was the main formation channel. 
Under this assumption, it has recently been suggested that we will be able to detect a
clustered signal of gravitational waves from merging black holes \citep{lake21}.
Globular clusters might contribute significantly to reionization \citep{ricotti2002,schaerer2011,mbk17,mbk18,ma21}, and it would be interesting to explore if a streaming 
velocity signal could be imprinted on the patchiness of reionization \citep{park21}. 

Translating the gaseous mass of our proto-cluster regions into masses of the resulting globular clusters strongly depends on the star formation efficiency employed. 
However, even for a SFE of 10\%, we find cluster masses between 1--8$\times 10^3$\,\Ms, significantly below the masses of classical globular clusters. As our simulations solely trace the gas, we can only speculate that any star clusters formed in streaming velocity regions might be $\sim 2$ orders of magnitude lower in stellar mass than typical globular clusters today ($M_{\star}=2\times 10^{5}$\,\Ms; \citealt{harris1991}) and therefore would dissolve relatively quickly due to two-body relaxation \citep{spitzer1987}.

We expect these proto-globular cluster regions to have luminosities between 
$4.5\times 10^{38}$\,erg\,s$^{-1}$ -- $2.2\times 10^{40}$\,erg\,s$^{-1}$ at formation 
when assuming the same model as \cite{chiou19}.\,\footnote{We take 
the starburst ``model A'' from \cite{schaerer03} that describes a top-heavy IMF. We further assume, again following \cite{chiou19}, a star formation efficiency of 10\% and a photon escape fraction of 50\%.} Such low-mass clusters forming at $z \gtrsim 10$ would thus have observed fluxes that are below the sensitivity limits of existing and upcoming telescopes, including the nJy sensitivity of the {\it James Webb Space Telescope} (JWST). 

We find that a metallicity of $Z_\mathrm{req} \sim 10^{-3}\,Z_\odot$ is required to cool proto-cluster regions to the CMB temperature floor.
This is 
roughly a factor of 3-5 lower than the observed metallicity floor of GCs in the Milky Way \cite[e.g.][2010 edition]{harris96}. 
However, dissolved, less massive GCs may have lower metallicities: the Phoenix stream likely consists of a single stellar population originating in a dissolved globular cluster of $Z = 10^{-2.7}Z_\odot$ \citep{wan20}. Another example, 
the Sylgr stream, might be the remnant of another disrupted GC with stars of an even lower metallicity, $Z = 10^{-2.9}Z_\odot$ \citep{roederer19}.
Larger surveys might reveal a whole population of these objects. 

A further intriguing scenario is one in which only a small subset of globular clusters form as a result of streaming velocities. These clusters would likely have different properties from the bulk of the GC population, as the physics of formation would differ substantially. The enigmatic galaxy NGC 1052-DF2 has an unusual population of globular clusters that are substantially more massive than typical GCs, and the galaxy itself has significantly less dark matter than is expected in the standard scenario if it has not undergone significant tidal interactions 
\citep{vanDokkum18,vanDokkum18b}. 
Could it be that the curious nature of NGC 1052-DF2 and its GCs have their origin in regions with significant $v_{\rm bc}$? Further work that includes the effects of metal cooling is necessary to understand any possible links.

\section{Conclusions and Outlook}
\label{sec:conclusions}
With our set of large hydrodynamical, high-resolution simulations, we have found that the first stars form inside dark matter halos at high redshift, 
even in regions of the Universe with a high streaming velocity of 2\,\s. 
A maximum number density of 100\,\cc\ provides a good threshold density for 
a high-density gas region to become Jeans unstable, hence enabling star formation. 
Gas overdensities with a maximum number density of less than 10\,\cc\ can 
be found outside the half-mass or virial radius of minihalos, 
primarily in regions of the Universe with a high streaming velocity. 
We find that primordial chemistry 
alone is not sufficient to cool these gas clouds down to low enough temperatures for 
collapse. 

More extensive studies are necessary to understand if 
streaming velocities provide a pathway to globular 
cluster formation, and if so, how important this formation channel is for the full GC population. 
If the globular clusters forming at high redshift are related to streaming 
velocities, we need improved simulations that include additional physical components. 
For one, the Pop~III star formation regime needs to be treated with 
the inclusion of Lyman-Werner radiation and star formation taking 
place. As Lyman-Werner radiation reduces the molecular hydrogen fraction 
in low-density regions, not only the number but also the morphology of the first star 
forming regions are altered. 
The study presented here provides an upper limit for the number of proto-globular clusters, 
as a smaller number of high density gas regions is expected 
when a Lyman-Werner background is considered.

We have shown in Section \ref{sec:metals} that metal-enrichment could 
cool gas outside of the halo half-mass radius to Jeans-unstable temperatures. 
A brief exploratory argument has shown that all gas regions, inside and outside the 
half-mass radius of a halo, can cool to the CMB temperature floor, and subsequently 
become Jeans unstable, when enriched to a metallicity of $Z\approx 10^{-3}Z_\odot$. 
Simulations that trace metal enrichment
from Pop~III supernova explosions \citep[such as e.g.][]{jeon14,chiaki19} 
need to be carried out with streaming velocities to understand 
if the metallicities required for Pop~II star formation can be reached 
at the right time. 
Ultimately, such simulations need to be continued to 
redshift $z=0$, to understand if globular clusters formed by the streaming velocity 
scenario can survive and 
reproduce the multiple stellar populations that are observed with the {\it Hubble Space Telescope} (HST).  

A major recent observational advance has been detections of proto-GC candidates in lensed HST observations coupled with VLT MUSE spectroscopy \citep{vanzella2017,vanzella2017a, vanzella2019},
and the very small sizes of many high-redshift objects 
\citep{kawamata2018,bouwens2021}
indicate that we may be on the cusp of routinely witnessing the birth of globular clusters in the distant Universe. It is very exciting that the next steps in this field with JWST will likely reveal proto-GCs in potentially large numbers at high redshifts 
\citep{carlberg2002,renzini2017, mbk18, pozzetti2019}. 
Indeed, JWST observations should provide crucial input on various unknown aspects of globular cluster formation such as the `mass budget' relating the mass of GCs at formation to their present-day mass \citep{mbk17,mbk18,pozzetti2019}. 
In the models considered so far, it has typically been assumed that globular clusters form predominantly at $10 \gtrsim z \gtrsim 3$. In the scenario where formation via the streaming velocity channel is important, these predictions will have to be substantially revised, as GC formation would likely proceed at higher redshifts that may be out of reach even for JWST. The stage is therefore set for important theoretical and observational advances in our understanding of globular cluster formation in the coming years.

\section*{Acknowledgments}
Support for this work was provided by NASA through the Hubble Fellowship grant HST-HF2-51418.001-A, awarded  by  STScI,  which  is  operated  by AURA,  under  contract NAS5-26555.
VB was supported by the National Science Foundation (NSF) grant AST-1413501. MBK acknowledges support from NSF CAREER award AST-1752913, NSF grant AST-1910346, NASA grant NNX17AG29G, and HST-AR-15006, HST-AR-15809, HST-GO-15658, HST-GO-15901, HST-GO-15902, HST-AR-16159, and HST-GO-16226 from STScI.
SCOG and RSK acknowledge funding from the European Research Council via the ERC Synergy Grant ``ECOGAL -- Understanding our Galactic ecosystem: From the disk of the Milky Way to the formation sites of stars and planets'' (project ID 855130). They also acknowledge support from the DFG via the Collaborative Research Center (SFB 881, Project-ID 138713538) ``The Milky Way System'' (sub-projects A1, B1, B2 and B8) and from the Heidelberg cluster of excellence (EXC 2181 - 390900948) ``STRUCTURES: A unifying approach to emergent phenomena in the physical world, mathematics, and complex data'', funded by the German Excellence Strategy.
The authors gratefully acknowledge the Gauss Center for Supercomputing for providing resources on SuperMUC at the Leibniz Supercomputing Center under project pr53ka.
\bibliography{refs}
\bibliographystyle{aasjournal}
%
\label{lastpage}
\end{document}